\documentclass[12pt]{article}
\usepackage{graphicx}
\usepackage{amsmath}
\usepackage{amsthm}
\usepackage{graphics}
\usepackage{epsfig}
\usepackage{amssymb}
\usepackage{xcolor}
 \usepackage{setspace}
\usepackage{extarrows}

\usepackage[authoryear]{natbib}
\usepackage{subfig}

\newcommand{\blind}{1}

\newcommand{\be}{\begin{eqnarray}}
\newcommand{\ee}{\end{eqnarray}}
\newcommand{\bea}{\begin{eqnarray*}}
\newcommand{\eea}{\end{eqnarray*}}

\newcommand{\E}{\mathbb{E}}

\newcommand{\kl}{\text{KL}}

\newtheorem{theorem}{Theorem}[section]

\newtheorem{example}{Example}[section]

\begin{document}
\def\spacingset#1{\renewcommand{\baselinestretch}%
{#1}\small\normalsize} \spacingset{1}

\if1\blind
{
  \title{\bf  \large Optimal Designs for Model Averaging in non-nested Models}
  \author{ \large Kira Alhorn, \hspace{.2cm}\\
   \large   Fakult\"at Statistik, Technische Universit\"at Dortmund\\
     \\
  \large    Holger Dette and Kirsten Schorning \thanks{
    The authors gratefully acknowledge financial support by the
Collaborative Research Center ``Statistical modeling of nonlinear
dynamic processes'' (SFB 823, Teilprojekt C2, T1) of the German Research Foundation
(DFG).} \\
 \large     Fakult\"at f\"ur Mathematik, Ruhr-Universit\"at Bochum}
  \maketitle
} \fi

\if0\blind
{
  \bigskip
  \bigskip
  \bigskip
  \begin{center}
    {\LARGE\bf Optimal Designs for Model Averaging in non-nested Models}
\end{center}
  \medskip
} \fi

\begin{abstract}
In this paper we construct optimal designs for frequentist model averaging estimation. We derive
the asymptotic distribution of the model averaging estimate with fixed weights  in the case where the competing models are non-nested
and none of these models is correctly specified. A Bayesian optimal design minimizes
an expectation  of the asymptotic  mean squared error
of the model averaging estimate  calculated with respect to a suitable prior distribution. 
We derive a necessary condition for the optimality  of a given with respect to this new criterion.
We demonstrate  that Bayesian optimal designs can improve the accuracy of  model averaging 
substantially. Moreover, the derived designs also improve  the accuracy of estimation in a model selected by model
selection and model averaging estimates with random weights.

\end{abstract}

\noindent%
\vfill

\newpage
\spacingset{1.5}
\section{Introduction}
\label{sec1}
\def\theequation{1.\arabic{equation}}
\setcounter{equation}{0}

There exists an enormous amount of literature on selecting an adequate model from a set of candidate models
for statistical analysis. Numerous model selection criteria have been developed for this purpose.  These procedures are widely used in practice and have the advantage of delivering a single model from a class of competing models, which makes them very attractive for practitioners. Exemplarily, we mention Akaike's information criterion (AIC), the
 Bayesian information   criterion  (BIC) and its extensions,    Mallow's $C_p$, the generalized cross-validation and   the minimum description length  (see the monographs of  \cite{burand2002}, \cite{konkit2008} and \cite{claeskens_model_2008} for more details).
 Different criteria have different  properties, such as consistency, efficiency and parsimony (used in the sense of \citet[Chapter 4]{claeskens_model_2008}).
 Overall there  seems  to be no universally optimal model selection criterion and different criteria might
   be preferable in different situations depending on the particular application.

On the other hand, there exists a well known post-selection problem in this approach because model selection introduces an additional variance that is often  ignored in statistical  inference after model selection (see \cite{poetscher_effects_1991} for one of the first contributions discussing this issue). This post-selection problem is inter alia attributable to the fact, that estimates after model selection behave like mixtures of potential estimates. For example, ignoring the model selection step (and thus the additional variability) may lead to confidence intervals  with coverage probability smaller than the nominal value,  see for example Chapter 7 in \cite{claeskens_model_2008} for a mathematical treatment of this phenomenon.

An alternative to model selection is model averaging, where estimates of a target  parameter are smoothed  across several models, rather than restricting inference on a single selected model.
This approach has been widely discussed in the Bayesian literature, where it   is known as  ``Bayesian model averaging'' (see the tutorial of \cite{hoeting1999} among many others).  For  Bayesian model averaging prior probabilities
 have to be specified. This might not always be  possible  and therefore
 \cite{hjort_frequentist_2003}  also proposed a    ``frequentist model averaging'', where smoothing across several models is commonly based on information criteria.
    \cite{kapetanios_forecasting_2008} demonstrated that  the frequentist approach
  is a worthwhile alternative to Bayesian model averaging.
 \cite{stock_forecasting_2003} observed  that averaging predictions usually performs better than forecasting in a single model. \cite{hong_bayesian_2012} substantiate these observations with theoretical findings for Bayesian model averaging
if the  competing models are ``sufficiently close''.  Further results pointing in this direction   can be found in
\cite{raft:2003},  \cite{schorning_model_2016} and \cite{buatois_2018}.

Independently of this discussion there  exists a large amount of research  how to optimally design experiments under model uncertainty
(see  \cite{boxhill1967,atkfed1975a} for early contributions). This work is motivated by the fact that  an optimal design
can improve the efficiency of the statistical analysis  substantially if the postulated model assumptions are correct, but may be inefficient if the model
is misspecified.  Many  authors  suggested   to choose the design for model discrimination such that  the power of a test  between competing regression models is maximized  (see \cite{ucibog2005,loptomtra2007,tomlop2010} or \cite{dette2015} for some more recent references).
Other authors  proposed to minimize an average of optimality criteria from different models
to obtain an efficient  design  for all  models under consideration (see    \cite{dette1990},  \cite{zentsa2002,Tommasi09}  among many others).

Although model selection  or averaging are commonly used tools for statistical inference under model uncertainty
most of the literature on designing experiments under model uncertainty does not address the specific aspects of these methods directly.
 Optimal designs are usually  constructed to maximize the power of a test for discriminating between competing models
or to minimize a functional of the asymptotic variance of estimates in the different models.
To the best of our knowledge   \cite{alhorn_optimal_2019}  is the first   contribution, which addresses the  specific challenges  of designing experiments
for  model selection or model  averaging.  These authors
 constructed optimal designs
 minimizing  the asymptotic mean squared error of the model averaging estimate and  showed
that optimal designs can yield  a reduction of the mean squared error  up to $45\%$.
Moreover, they also showed  that these designs improve the performance of estimates in models chosen by model selection criteria.
However, their theory relies heavily on the assumption of   nested models embedded in a framework of local alternatives
as developed by   \cite{hjort_frequentist_2003}.

The goal of the present contribution is the construction of optimal designs for  model averaging in cases where the competing models are not nested (note that in this case local alternatives cannot be formulated).
Moreover,  in contrast to most of the literature, we also consider the situation where all  competing models  misspecify the data underlying truth.
In order to derive  an optimality criterion, which can be used for the determination of optimal  designs  in this context, we
further  develop the approach of \cite{hjort_frequentist_2003} and derive
an asymptotic theory for model averaging estimates for classes of competing models which are non-nested.
Optimal designs are then constructed minimizing the asymptotic mean squared
error of the model averaging estimate and  it is demonstrated that  these designs yield substantially more precise model
averaging estimates. Moreover, these designs also improve the performance of  estimates after model selection. 
Our work also contributes to the discussion of  the superiority  of model averaging over  model selection. Most of the
results presented in literature indicate that  model averaging has some advantages over model selection in general. We demonstrate that
conclusions of this type depend sensitively on the class of models under consideration. In particular  we observe some advantages
of estimation after model selection if the competing models are of rather different shape. Nevertheless, the optimal designs developed
in this paper improve both estimation methods, where the improvement can be substantial in many cases.

The remaining part of this paper is organized as follows. The pros and cons  of model averaging and
model selection are
briefly discussed
in Section \ref{sec2} where we introduce the basic methodology  and 
 investigate the impact of  similarity of the candidate models  on the
performance of the  different estimates.  In Section \ref{sec3} we develop
asymptotic theory  for  model averaging estimation in the case where the models are non-nested and
all competing models might misspecify the underlying truth.
 Based on these results we derive a criterion for the determination of   optimal designs {and establish a necessary condition for optimality}. In   Section \ref{sec5}  we study
 the performance of these designs by means of  a simulation study.
 Finally, 
  technical  assumptions and proofs  are given
  Section \ref{app}.

\section{Model averaging versus model selection}
\label{sec2}
\def\theequation{2.\arabic{equation}}
\setcounter{equation}{0}

In this section we  introduce the basic terminology and  also  illustrate in a regression framework
that the superiority  of  model averaging about  estimation in a model chosen by  model selection  depends sensitively
on the class of competing models.

\subsection{Basic terminology}  \label{sec21}

We consider data obtained at $k$ different experimental conditions, say $x_1,\ldots,x_k$ chosen in a design space $\mathcal{X}$.
At   each experimental condition $x_i$ one  observes $n_i$ responses, say $y_{i1},\ldots,y_{in_i}$ $(i=1,\ldots,k)$, and the   total sample size is
$n=\sum_{i=1}^k n_i$. We also assume that  the responses $y_{i1},\ldots,y_{in_i}$ are realizations of random variables  of the form
\begin{align} \label{regression_model}
	Y_{ij} = \eta_s(x_i,\vartheta_s) + \varepsilon_{ij}, i=1,\ldots,k, j=1,\ldots,n_i, s=1,\ldots,r,
\end{align}
where the regression function $\eta_s$ is a differentiable function with respect to the parameter $\vartheta_s $ and the random errors $\varepsilon_{ij}$ are independent normally distributed with mean 0 and common variance $\sigma^2$.  Furthermore, the index $s$ in  $\eta_s$   corresponds to  different models (with parameters   $\vartheta_s $) and we assume that there are
$r$ competing regression functions  $\eta_{1}, \ldots , \eta_{r}$ under  consideration.

Having $r$ different candidate models (differing by the regression functions $\eta_s$) a classical approach for estimating a parameter of interest, say $\mu$, is to calculate an information criterion for each model under consideration
 and estimate this parameter in the model optimizing this criterion. For this purpose,
we  denote the density of the normal distribution corresponding to a regression model \eqref{regression_model}
by \mbox{$f_s(~\cdot  \mid x_i,\theta_s)$} with parameter $\theta_s = (\sigma^2,\vartheta_s)^{\top}$
and identify the different models by their densities  $f_{1}, \ldots , f_{r}$ (note  that in the situation considered in this sections
these only differ in the mean).
Using the observations $y_n=(y_{11},\ldots,y_{1n_1},y_{21},\ldots,y_{kn_k})^{\top}$ 
 we calculate in each model the  maximum likelihood estimate
 \begin{align} \label{quasiloglikest}
	\hat{\theta}_{n,s} = \text{arg} \max_{\theta_s \in \Theta_s} \ell_{n,s}(\theta_s  \mid y_n) 
\end{align}
   of the parameter $\theta_s$, where
   \begin{align} \label{quasiloglik}
	\ell_{n,s}(\theta_s  \mid y_n) = \tfrac{1}{n} \sum_{i=1}^k \sum_{j=1}^{n_i} \log f_s(y_{ij}  \mid x_i,\theta_s)
\end{align}
is  the  log-likelihood in candidate model $f_{s}$ ($s=1, \ldots r$).
Note, that we do not assume that  the true data generating density is  included in the set of candidate models
$f_1, \ldots  , f_r$.   Each   estimate $\hat{\theta}_{n,s} $  of the parameter $\theta_s$ yields
an estimate
	$\hat{\mu}_s = \mu_s(\hat{\theta}_{n,s})$
for the quantity of interest,   where $\mu_s$ is the target parameter  in model $s$.

For example, regression models   of the   type \eqref{regression_model} are frequently used in dose finding studies
(see \cite{ting_dose_2006} or  \cite{bretz_dose_2008}).
In this case  a typical  target function $\mu_s$ of interest  is the ``quantile'' defined by
 \begin{equation} \label{EDPgen}
\mu_{s}(\theta_s)  = \inf \left\{ x \in {\cal X }~ \Big |~\tfrac{ \eta_s (x, \vartheta_s) - \eta_s (a, \vartheta_s) }{\eta_s (b, \vartheta_s) - \eta_s (a, \vartheta_s)} \geq \alpha   \right\} ~.
\end{equation}
The value defined in \eqref{EDPgen} is well-known as $\text{ED}_\alpha$, that is, the effective dose at which $100\times \alpha \%$  of the maximum effect in the design space ${\cal X }= [a,b]$
is achieved.

We now briefly discuss the principle of model selection and averaging  to estimate the target parameter $\mu$.
For model selection   we choose the   model $f_{s^*}$ from $f_1, \ldots , f_s $, which   maximizes  Akaike's   information criterion (AIC)
\begin{align} \label{aic_def}
  \text{AIC}(f_s  \mid y_n) = 2\ell_{n,s}(\hat{\theta}_{n,s} \mid y_n) - 2 p_s ,
\end{align}
where  $ p_s$ is the number of parameters in model $f_s$  (see \cite{claeskens_model_2008},~Chapter 2).
The target parameter is finally estimated by $\hat \mu = \mu_{s^*}(\hat{\theta}_{n,{s^*}})$.
Obviously,  other  model selection schemes, such as the Bayesian or focussed information criterion can be used here as well, but
we restrict ourselves to the AIC for the sake of a transparent presentation.

Roughly speaking,    model averaging
 is a weighted average of the individual estimates in the competing models.  It  might be viewed from a Bayesian (see for example \cite{wass2000}) or a frequentist point of view (see for example \cite{claeskens_model_2008}) resulting in different choices of model averaging weights. We will focus here on non-Bayesian methods.
More explicitly, assigning nonnegative weights $w_1,\ldots ,w_r$  to the   candidate models  $f_1,\ldots ,f_r,$ with $\sum_{i=1}^rw_i=1$, 
the model averaging estimate for $\mu$ is given by
\begin{align} \label{mu_mav}
	\hat{\mu}_{\text{mav}} = \sum_{s=1}^r w_s \mu_s(\hat{\theta}_{n,s}).
\end{align}
Frequently  used   weights    are uniform weights (see, for example  \cite{stock_combination_2004},  \cite{kapetanios_forecasting_2008}). More elaborate model averaging weights can be chosen depending on the data. For example, \cite{claeskens_model_2008} define  smooth AIC-weights as
  \begin{equation}\label{smooth_aic_weights}
w^{{\footnotesize \text{smAIC}}}_{s} (y_n) = \tfrac{\exp\{\tfrac{1}{2} \text{AIC}(f_s \mid y_n)\}}{\sum_{s=1}^{r} \exp\{\tfrac{1}{2} \text{AIC}(f_s \mid y_n)\}}.
\end{equation}
Alternative   data dependent weights can be constructed using other information criteria or model selection criteria. There
also exists a vast amount of literature on determining optimal data dependent weights such that the resulting mean squared error of the model averaging estimate is minimal (see \cite{hjort_frequentist_2003}, \cite{hansen_least_2007} or \cite{liang_optimal_2011} among many others). For the sake of brevity  we concentrate
 on smooth AIC-weights here, but  similar observations  as presented in this paper can also be made for other data dependent weights.

  \subsection{The class of competing models matters}
  \label{sec22}

In this section we  illustrate the influence of the candidate set on
the  properties of   model averaging estimation and estimation after model selection
by means of   a brief simulation study.
For this purpose  we   consider four  regression models
of the form \eqref{regression_model}, which are commonly used in dose-response modeling
and  specified in  Table \ref{tab1} with corresponding parameters.

   \begin{table}[t]
 \centering
 \renewcommand{\arraystretch}{.9}
 \small
 \begin{tabular}{| c | c | c|}
 \hline
Model & Mean function  $\eta _s$ & Parameter specifications \\ \hline
Log-Linear ($f_{1}$) & $\eta_1(x_i,\vartheta_1) = \vartheta_{11} + \vartheta_{12} \log (x_i + \vartheta_{13})$ & $\vartheta_1 = (0,0.0797,1)^{\top}$ \\
Emax  ($f_{2}$) & $\eta_2(x_i,\vartheta_2) = \vartheta_{21} + \tfrac{\vartheta_{22}x}{\vartheta_{23} + x}$ & $\vartheta_2 = (0,0.467,25)^{\top}$ \\
Exponential  ($f_{3}$)  & $\eta_3(x_i,\vartheta_3) = \vartheta_{31} + \vartheta_{32} \exp (x_i/\vartheta_{33})$ & $\vartheta_3 = (-0.08265,0.08265,85)^{\top}$ \\
Quadratic   ($f_{4}$) & $\eta_4(x_i,\vartheta_4) = \vartheta_{41} + \vartheta_{42} x + \vartheta_{43} x^2$ & $\vartheta_4 = (0, 0.00533, -0.00002)^{\top}$ \\
  \hline
 \end{tabular}
 \renewcommand{\arraystretch}{1}
 \caption{\it
 Models and parameters used for the simulation study.}
\label{tab1}
 \end{table}
Here we adapt the setting of \cite{pinheiro_design_2006} who model the dose-response relationship of an anti-anxiety drug, where the dose of the drug may vary in the interval ${\cal X} = [0,150]$. In particular, we have  $k=6$ different dose levels $x_i \in \{0, 10, 25, 50, 100, 150\}$
and patients are allocated  to each dose level most equally, where the total sample size is  $n \in \{50,100,250\}$.
 We   consider the problem
 of estimating the $\text{ED}_{0.4}$, as defined in \eqref{EDPgen}.

\begin{figure}[t]
\centering
\subfloat{\includegraphics[width=0.45\textwidth]{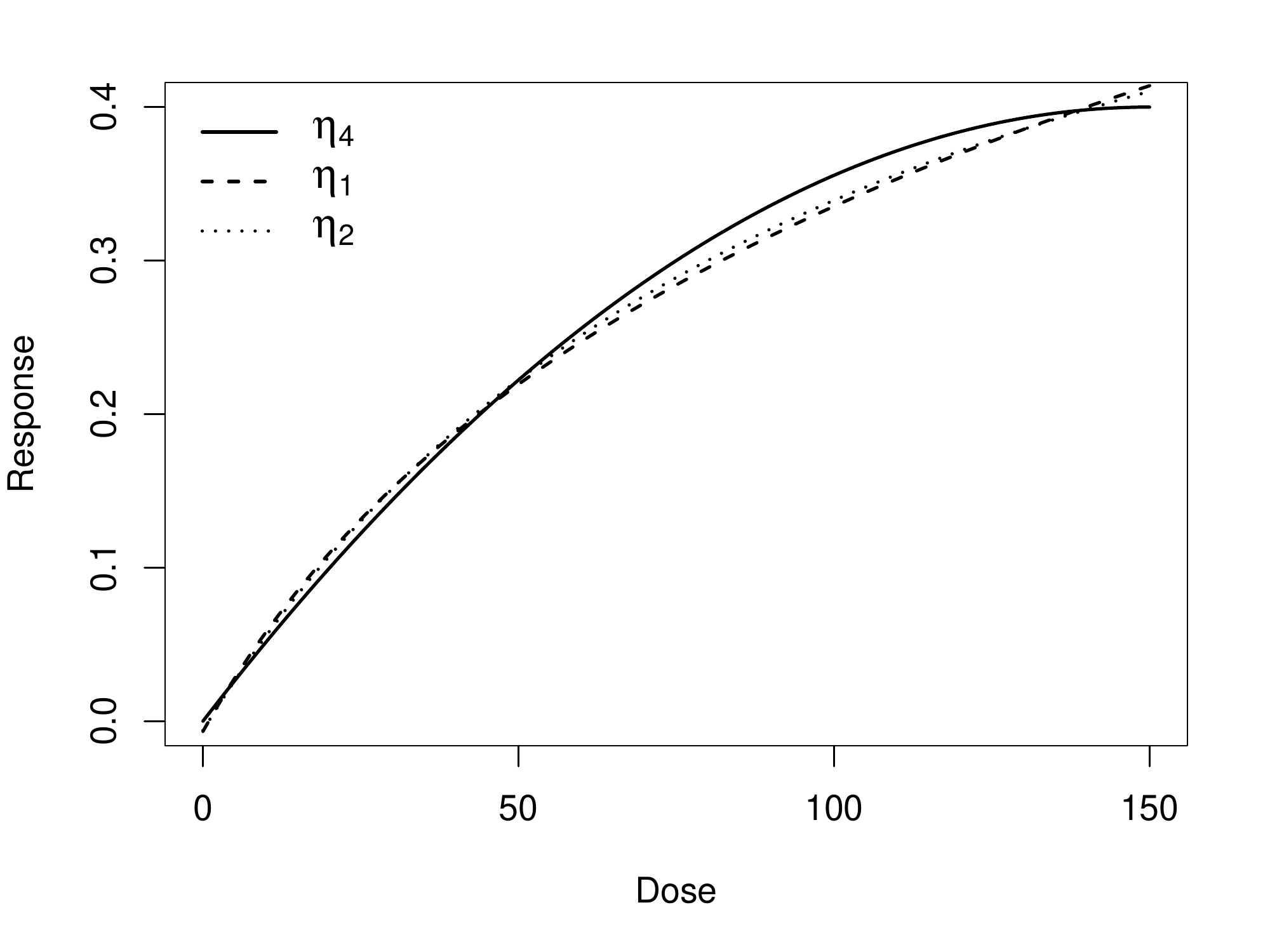}
} \qquad
\subfloat{\includegraphics[width=0.45\textwidth]{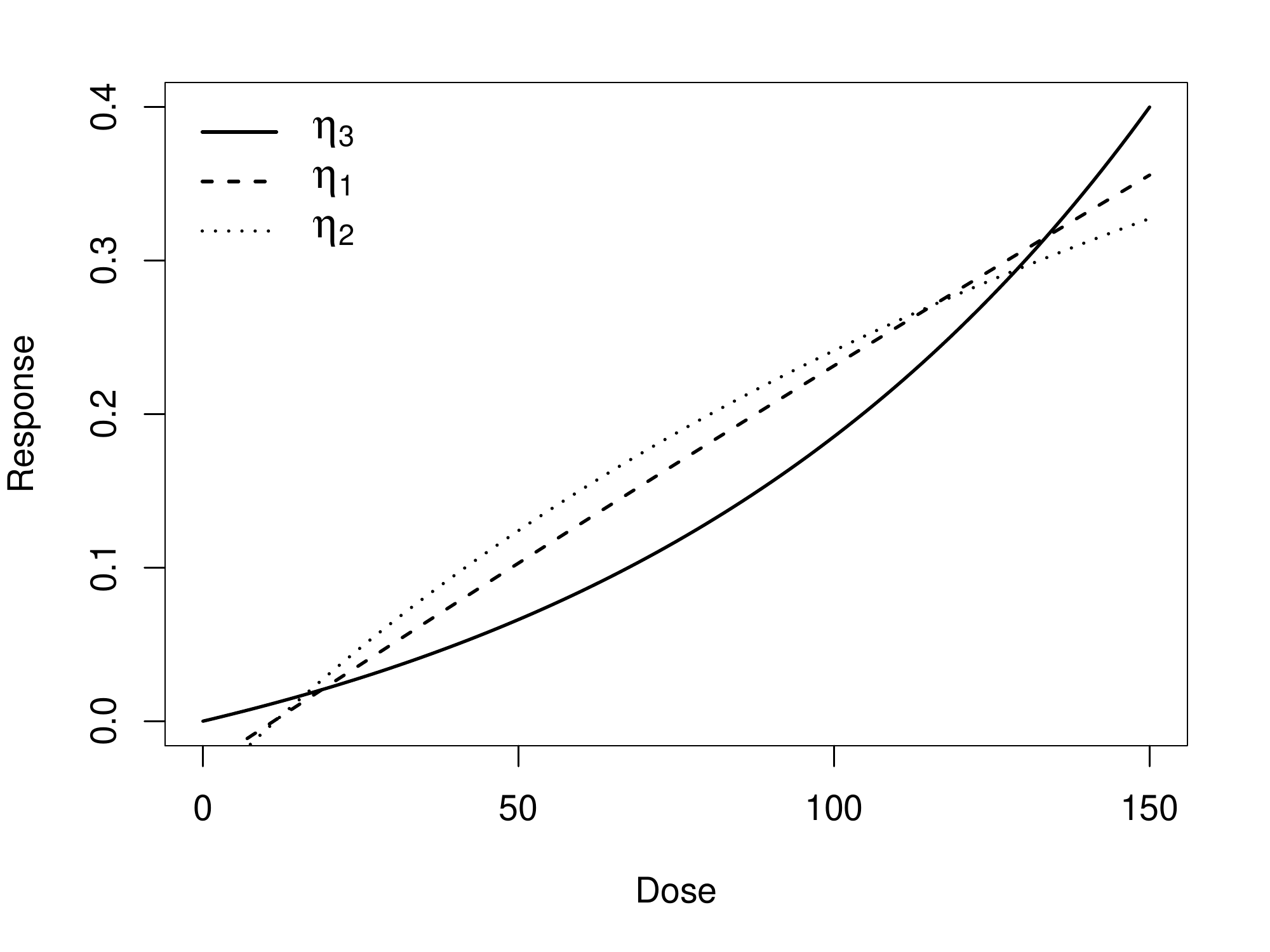}
} \\
\caption{\it Left panel: quadratic  model  (solid line) and its best approximations
 by the log-linear (dashed line) and  the Emax model (dotted line) with respect to the Kullback-Leibler divergence
 \eqref{klsim}.  Right panel: exponential  model   (solid line) and its best approximations
 by the log-linear (dashed line) and  the Emax model (dotted line).}
\label{fig1}
\end{figure}

 To investigate the particular differences between both estimation methods  we choose two different sets of competing models from Table \ref{tab1}.
 The first  set
 \begin{equation}
 \label{set1}
 {\cal S}_1=
 \{ f_1 , f_2, f_4 \}
 \end{equation}
   contains  the log-linear, the Emax and the quadratic model, while the second set
  \begin{equation}
 \label{set2}
 {\cal S}_2=
 \{ f_1 , f_2, f_3 \}
 \end{equation}
contains the log-linear, the Emax and the exponential model. The set ${\cal S}_1$ serves as a prototype  set  of  ``similar'' models
while the set  ${\cal S}_2$ contains models of  more  ``different'' shape.  This
is   illustrated in Figure \ref{fig1}. In the left panel
we show  the quadratic model  $f_4$  (for the parameters specified in Table \ref{tab1})
and the best approximations  of this function
by   a log-linear model ($f_1$)    and an Emax model  ($f_2$) with respect to the
Kullback-Leibler divergence
\begin{equation}  \label{klsim}
	  \tfrac{1}{6 } \sum_{i=1}^6  \int f_4(y  \mid x_i, \theta_4) \log \left( \tfrac{f_4(y  \mid x_i,\theta_4)}{f_s(y  \mid x_i ,\theta_s)} \right) dy~ ,~~s=1,2.
\end{equation}
In this case, all models have a very similar shape and we obtain for the ED$_{0.4}$ the values $32.581$, $32.261$ and $33.810$
for the log-linear ($f_{1}$), Emax ($f_{2}$) and quadratic model ($f_{4}$).
Similarly  the   right panel shows the   exponential model ($f_{3}$, solid line) and its corresponding best approximations
by the log-linear model ($f_1$) and the Emax model ($f_2$).  Here we observe larger differences between the models in the candidate
set and  we obtain for the ED$_{0.4}$ the values $58.116$, $42.857$ and $91.547$
for  the models $f_{1}$,  $f_{2}$ and $f_{3}$, respectively.

All results presented  in this paper  are based on $1000$ simulations runs generating in each run $n$ observations
of the form
\begin{equation}\label{sim_values}
y_{ij}^{(l)} = \eta_s(x_i,\vartheta_s) +  \varepsilon_{ij}^{(l)}
,i=1,\ldots,k,j=1,\ldots,n_i,
\end{equation}
 where  the errors    $\varepsilon_{ij}^{(l)}$  are independent centered  normal distributed random variables
 with $\sigma^2=0.1$ and $\eta_s $  is one of the models $\eta_{1}, \ldots , \eta_{4}$ (with  parameters
 specified in Table \ref{tab1}).   The  parameter $\mu= \text{ED}_{0.4}$ is estimated
 by model averaging  with  uniform weights, smooth AIC-weights given in  \eqref{smooth_aic_weights} and  estimation after model selection by the AIC criterion.

In Table \ref{tab2} and  \ref{tab3} we   show the simulated mean squared errors
of the  model averaging estimates
with uniform weights (left column),   smooth AIC-weights (middle column) and
 estimation after model selection (right column).
 Here, different rows correspond to different models.
  The numbers printed in bold face indicate the estimation method with the smallest mean squared error.

\subsubsection{Models of similar shape} \label{sec221}

We will first  discuss the results for  the  set  of similar models    in  \eqref{set1} (see Table \ref{tab2}).  If the data generating model is an element of  the set
of candidate models, model averaging  with uniform  weights performs very well.
Model averaging with smooth AIC-weights  yields an about  $10\%$ -$25\%$ larger mean squared error
(except for two cases, where it performs better than model averaging with uniform  weights).
On the other hand the mean squared error of estimation  after   model selection is  substantially larger than that
of  model averaging, if the sample size is small.  This   is a consequence  of the additional variability associated with data-dependent weights.
For example, if the sample size is  $n=50$ and  the data generating model is given by $f_{1}$, the mean squared errors
of the model averaging estimates with  uniform  and    smooth AIC-weights  and   the estimate after   model selection
are given by  $437.0$,  $498.3$ and  $759.0$, respectively.
The corresponding variances are given by $235.2$, $337.6$ and $599.7$, respectively.  For the squared bias
the order is exactly the opposite, that is  $ 201.9$,  $160.7$, $159.3$, but the differences are not so large.
This means that the bias can be reduced by using random weights, because these put more weight on the ``correct'' model.
As a consequence, compared to model averaging with uniform weights  the performance of  model averaging with
smooth AIC-weights  and   the estimate after   model selection improves with increasing sample size.
Nevertheless, if the ``true'' model is an element of the candidate set and the functions in this set have a similar shape,
model averaging  performs better than estimation after  model selection. In particular, model averaging with (fixed) uniform weights
yields very reasonable results. These observations coincide with the findings of
\cite{schorning_model_2016}  and \cite{buatois_2018} who  compared model averaging
 and  model selection  in the context of dose finding studies (see  also \cite{longmei_model_2018} 
 for similar results   for the AIC in the context of ordered probit and nested logit models).

\begin{table}[t]
\centering
 \renewcommand{\arraystretch}{.9}
 \small
\begin{tabular}{| c | c || c | c | c |}
 \hline
\cline{3-5}
 model & sample size & uniform  weights &  smooth AIC-weights & model selection \\
 \hline
 & $n=50$ & \textbf{437.045} & 498.323 & 758.978 \\
 $f_1$  & $n=100$ & 223.291 & \textbf{218.99} & 285.062 \\
  & $n=250$ & 111.973 & 82.713 & \textbf{78.371} \\ \hline
 & $n=50$ & \textbf{286.638} & 329.904 & 515.32 \\
  $f_2$  & $n=100$ & \textbf{189.785} & 203.796 & 251.836 \\
  & $n=250$ & \textbf{62.792} & 64.854 & 66.54 \\ \hline
& $n=50$ & \textbf{276.037} & 361.101 & 669.873 \\
 $f_4$   & $n=100$ & \textbf{190.662} & 244.558 & 391.443 \\
  & $n=250$ & \textbf{92.653} & 109.852 & 139.859 \\
   \hline
   \hline
 & $n=50$ & 1503.903 & \textbf{1372.31} & 1381.033 \\
    \textbf{$f_3$}  &  $n=100$ & 1109.622 & 856.484 & \textbf{729.912} \\
  & $n=250$ & 864.163 & 398.144 & \textbf{255.604}  \\ \hline

\end{tabular}
\caption{\it  Simulated mean squared error of different  estimates of the $\text{ED}_{0.4}$. The set of candidate models is
${\cal S}_{1}= \{f_1,f_2,f_4\}$.
Left  column: model averaging    with uniform weights; middle column: model averaging     with  smooth AIC-weights;
right column: estimation after  model selection. }
 \renewcommand{\arraystretch}{1}
\label{tab2}
\end{table}

The situation changes if  none of the candidate models from the set ${\cal S}_{1}$
is the ``true'' model.  This is illustrated in the lower part of Table \ref{tab2}, where
we show results if  the exponential model $f_3$ is used for generating the data.  We observe that model averaging  with uniform
weights is outperformed by model averaging   with  smooth AIC-weights. Moreover, the estimate after  model selection
is even better, if the sample size increases. These observations can be explained by  the different shapes of the regression functions, as illustrated in Figure \ref{fig1}. By a suitable choice of parameters  the quadratic model can adapt to the shape of the exponential model, whereas the log-linear and the Emax model still have different forms (see right panel of Figure \ref{fig1} for the best approximations that are possible using the log-linear and the Emax model).
Thus, incorporating these models in a model averaging estimate yields a large bias, that can be reduced substantially by
data dependent weights or by model selection.  For example, if $n=100$  the squared bias of the model averaging estimate with  uniform  weights
is $981.631$, whereas the model averaging estimate with   smooth AIC-weights and the estimate after model selection show a squared bias of $328.634$ and $69.465$, respectively.

\subsubsection{Models of more different  shape} \label{sec222}

We will now consider the candidate set ${\cal S}_{2}$ in \eqref{set2}, which serves as an example of  more different models
and includes  the log-linear, the Emax and the exponential model.
The  simulated mean squared errors of the three estimates of the ED$_{0.4}$ are given in Table \ref{tab3}.
The upper part of the table corresponds to cases, where   data  is  generated from a model in the candidate set ${\cal S}_{2}$ used for model selection and averaging.
  In contrast to Section \ref{sec221}  we observe only one scenario, where model averaging with  uniform weights gives  the smallest mean squared error  (but in this case  model averaging with smooth AIC-weights yields very similar results).  If the sample size
  increases  model averaging with smooth AIC-weights  and estimation after  model selection
  yield  a substantially smaller mean squared error.  An explanation of this observation
  consists in the fact that for a candidate set containing models with a  rather different shape
  model averaging  with uniform weights   produces a large bias. On the other hand
  model averaging with smooth AIC-weights  and estimation after  model selection
  adapt  to the data and put more weight on the ``true'' model, in particular if the sample size is large.
As estimation after model selection has a larger variance  and  the variance is  decreasing with increasing sample size,
 the bias is dominating the mean squared error for large sample sizes and thus estimation in the model selected by the AIC
 is more efficient for large sample sizes.

\begin{table}[t]
\centering
 \renewcommand{\arraystretch}{.9}
 \small
\begin{tabular}{| c | c || c | c | c |}
 \hline
 & & \multicolumn{3}{|c|}{estimation method} \\
\cline{3-5}
 model & sample size & uniform  weights &  smooth AIC-weights & model selection \\
 \hline
 & $n=50$ & 834.295 & \textbf{553.427} & 776.311 \\
$f_1$   & $n=100$  & 712.404 & \textbf{340.254} & 353.707 \\
   &$n=250$ & 524.518 & 48.587 & \textbf{38.591} \\  \hline
 & $n=50$ & 640.706 & \textbf{505.054} & 669.285 \\
  $f_2$  & $n=100$ & 517.963  & \textbf{267.967}   & 286.272 \\
  & $n=250$ & 394.536 & 65.805 & \textbf{53.424} \\  \hline
& $n=50$ & \textbf{1076.154} & 1141.476 & 1427.441 \\
  $f_3$    & $n=100$ & 871.362 & \textbf{766.140} & 802.763 \\
  & $n=250$& 802.196 & 480.641 & \textbf{399.839} \\ \hline
  \hline
 & $n=50$ & \textbf{288.091} & 486.501 & 852.377 \\
   $f_4$ & $n=100$ & \textbf{208.628} & 298.315 & 419.651 \\
  & $n=250$& 162.689 & \textbf{138.331} & 142.673 \\
   \hline
\end{tabular}
\caption{ \it  Simulated mean squared error of different  estimates of the $\text{ED}_{0.4}$. The set of candidate models is
${\cal S}_{2}= \{f_1,f_2,f_3\}$.
Left  column: model averaging   with uniform weights; middle column: model averaging    with  smooth AIC-weights;
right column: estimation after  model selection.}
 \renewcommand{\arraystretch}{1}
\label{tab3}
\end{table} 

   Finally, if the data is generated according to   the quadratic model $f_4 \not \in {\cal S}_{2}$
   model averaging with  uniform weights has the smallest mean squared error if the sample is $n=50$ and $n=100$. In this case
   estimation in the model selected by the AIC performs much worse (due to its large variance).  However,
   the differences become smaller with increasing sample size. In particular for $n=250$ model averaging with smooth AIC-weights and
   estimation after  model selection show a substantially better performance than model averaging with uniform weights.

\medskip

The numerical  study in Section \ref{sec221} and \ref{sec222} can be summarized as follows. The results observed in the literature have to be partially  relativized.
The superiority of model averaging with  uniform weights can only be observed for classes of ``similar'' competing models and a not too large signal to noise
ratio. On the other hand if the models in the candidate set are of rather different structure, model averaging with data dependent weights (such as smooth AIC-weights)
or estimation after  model selection may show a better performance.  For these reasons we will investigate optimal/efficient designs for
all three estimation  methods in the following sections. We will demonstrate that a careful design of experiments can improve the accuracy of
these estimates substantially.

\section{Asymptotic properties and optimal design} \label{sec3}
\def\theequation{3.\arabic{equation}}
\setcounter{equation}{0}

In this section we will derive the asymptotic properties  of model averaging estimates with fixed  weights in the case where the competing models are not nested. The results can  be used for (at least) two purposes.
On the one  hand they provide some understanding of  the empirical findings in  Section \ref{sec2}, where  we observed, that for increasing sample size
the mean squared error of model averaging estimates is dominated by its bias. On the other hand, we will use these results to develop an asymptotic representation of the mean squared error of the
model averaging estimate, which can be used in the construction of optimal designs.

\subsection{Model averaging for non-nested models} \label{sec31}

\cite{hjort_frequentist_2003} provide an asymptotic distribution of frequentist model averaging estimates making  use of local alternatives  which require the true data generating process to lie inside a wide parametric model. All candidate models are sub-models of this wide model and the deviations in the parameters are restricted to be of order $n^{-1/2}$. Using this assumption results in
convenient approximations for the mean squared error as  variance and bias are both of order $O(1/n)$.    However, in the discussion of this paper
  \cite{raft:2003}  pose the question if  the framework of local alternatives is
realistic. More importantly,
frequentist model averaging is also often used for non-nested models  (see for example \cite{verrier_dosefinding_2014}).
In this section we will develop  asymptotic theory for model averaging estimation
in non-nested models. In particular, we do not assume
that the ``true''  model is among  the  candidate models used in the  model averaging estimate.

As we will apply  our  results for the construction of efficient designs for model averaging estimation
we use the common  notation of this field.
To be precise, let $Y$ denote a response variable and let $x$ denote a vector of explanatory variables
defined on a given compact design space ${\cal X}$. Suppose  that $ Y$ has   a density  $g (y  \mid  x) $
with respect to a dominating measure. For estimating a quantity of interest, say $\mu$,  from the distribution $g$ we use $r$ different parametric candidate models with  densities
\begin{equation} \label{candmod}
f_1( y \mid x,\theta_1),\ldots,f_r( y \mid x ,\theta_r)
\end{equation}
where   $\theta_s $  denotes the parameter in the $s$th model, which  varies in a compact parameter space, say $\Theta_s \subset \mathbb{R}^{p_s}$
 $(s=1,...,r)$.  Note, that in general
 we do not assume that the density $g$ is contained in the set of candidate models  in \eqref{candmod} and that the regression
 model \eqref{regression_model}  investigated in  Section \ref{sec2}
 is a special case of this  general notation.

 We  assume  that $k$ different experimental conditions, say $x_1,\ldots,x_k$, can be chosen in a design space $\mathcal{X}$ and that at each experimental condition $x_i$ one can observe $n_i$ responses, say $y_{i1},\ldots,y_{in_i}$
 (thus the total sample size is $n=\sum_{i=1}^k n_i$), which
 are  realizations of independent identically  distributed random variables $Y_{i1},\ldots,Y_{in_i}$ with density
  $g(\cdot  \mid  x_{i})$.   For example, if $g$ coincides with $f_{s}$ then the density of the random variables
 $Y_{i1},\ldots,Y_{in_i}$ is given by  $f_s(~\cdot  \mid x_{i} ,\theta_s) $ ($i=1,\ldots , k$).
To measure  efficiency and to compare different experimental designs we will use  asymptotic arguments
 and  consider the case  $\lim_{n \to \infty} \tfrac{n_{i}}{n}=\xi_{i} \in (0,1)$ for $i=1, \ldots, k$.
 As common in optimal  design theory  we collect this information  in the form
\begin{equation} \label{design}
\xi= \left \{
x_{1} ,  \ldots  , x_k ;
\xi_{1} ,  \dots ,  \xi_{k}
\right\},
\end{equation}
which is called approximate design in the following discussion (see, for example,  \cite{kiefer_general_1974}).
For  an approximate design $\xi$  of the form \eqref{design} and total sample size   $n$ a rounding procedure is applied to obtain integers $n_{i}$ taken at each $x_i$ ($i=1,\ldots,{k})$
from the not necessarily integer valued quantities $\xi_{i}n$  (see, for example \cite{pukelsheim_optimal_2006}, Chapter 12).

The asymptotic properties of the  maximum likelihood estimate 
(calculated under the assumption that $f_{s}$ is the correct density) is derived under certain  assumptions of regularity  (see the  Assumptions (A1)-(A6) in Section \ref{app}).  In particular, we assume
  that  the functions $f_s$  are twice  continuously differentiable with respect to
   $\theta_s$ and that  several  expectations of  derivatives of the log-densities exist.
For a  given  approximate design $\xi$ and a candidate density $f_s$ we denote  by
\begin{align} \label{kldist}
	\kl(g:f_s \mid \theta_s,\xi) = \int g(y  \mid x) \log \left( \tfrac{g(y  \mid x)}{f_s(y  \mid x ,\theta_s)} \right) dy d\xi(x),
\end{align}
 the Kullback-Leibler divergence  between the models $g$ and $f_{s}$
and assume that
	\begin{align} \label{kl_min}
		\theta_{s,g}^*(\xi) = \text{arg}\min_{\theta_s \in \Theta_s} \kl(g:f_s \mid \theta_s,\xi)
	\end{align}
is unique	 for each $s\in\{1,\ldots,r\}$.    For notational simplicity we will omit the dependency of the minimum on the density $g$, whenever
	it is clear from the context and denote the minimizer by $\theta_{s}^*(\xi)$.
	We also  assume  that the
	 matrices
\begin{align}
	A_s(\theta_s,\xi) &= \sum_{i=1}^{k }\xi_{i} \  \E_{g(\cdot  \mid x_{i} )} \Big ( \tfrac{\partial^2 \log f_s(Y_{ij} \mid x_i,\theta_s)}{\partial \theta_s \partial \theta_s^{\top}} \Big )
	,\label{As} \\
	B_{st}(\theta_s,\theta_t,\xi) &= \sum_{i=1}^{k }\xi_{i} \   \E_{g(\cdot  \mid x_{i} )}
	 \Big ( \tfrac{\partial \log f_s(Y_{ij} \mid x_i,\theta_s)}{\partial \theta_s} \Big ( \tfrac{\partial \log f_t(Y_{ij} \mid x_i,\theta_t)}{\partial \theta_t} \Big )^{\top} \Big )
	 , \label{Bst}
\end{align}
exist,  where expectations are taken with respect to the true distribution ${g(\cdot  \mid x_{i} )}$.

Under standard assumptions \cite{white_1982} shows the existence of a measurable  maximum likelihood estimate $\hat{\theta}_{n,s}$ for all candidate models
which is strongly consistent
for the (unique) minimizer $\theta_s^*(\xi)$   in \eqref{kl_min}. Moreover,
the estimate is also asymptotically  normal  distributed, that  is
\begin{align} \label{as_norm}
	\sqrt{n}(\hat{\theta}_{n,s} - \theta_s^*(\xi)) \overset{\mathcal{D}}{\longrightarrow} \mathcal{N}\left(0,A_s^{-1}(\theta_s^*(\xi))B_{ss}(\theta_s^*(\xi),\theta_s^*(\xi))A_s^{-1}(\theta_s^*(\xi))\right),
\end{align}
where we assume the existence of the inverse matrices, $\overset{\mathcal{D}}{\longrightarrow}$ denotes convergence in distribution and we use  the notations
\begin{equation} \label{asnot}
A_s(\theta_s^*(\xi)) = A_s(\theta_s^*(\xi), \xi ) ~ , ~~B_{st}(\theta_s^*(\xi),\theta_t^*(\xi)) = B_{st}(\theta_s^*(\xi),\theta_t^*(\xi), \xi)
\end{equation}
($s,t=1, \ldots r$). The following result
gives  the asymptotic distribution of model averaging estimates of the form \eqref{mu_mav}.

\begin{theorem} \label{theorem:distrmu} 
If Assumptions (A1) - (A7) in Section \ref{app1} are satisfied, then
	the model averaging estimate \eqref{mu_mav} satisfies
	\begin{align} \label{distr_mumav}
		\sqrt{n} \Big (\hat{\mu}_{\text{mav}} - \sum_{s=1}^r w_s \mu_s(\theta_s^*(\xi))\Big ) \overset{\mathcal{D}}{\longrightarrow} \mathcal{N}
		\left(0, \sigma_w^2(\theta^*(\xi))
		\right),
	\end{align}
	where the asymptotic variance is given by
	\begin{align} \label{sigma_w}
		\sigma_w^2(\theta^*(\xi)) = \sum_{s,t=1}^r  w_s w_t \Big ( \tfrac{\partial \mu_s (\theta_s^*(\xi))}{\partial \theta_s} \Big )^{\top} A_s^{-1} (\theta_s^*(\xi)) B_{st} \left(\theta_s^*(\xi),\theta_t^*(\xi)\right) A_t^{-1} \left(\theta_t^*(\xi)\right) \tfrac{\partial \mu_t (\theta_t^*(\xi))}{\partial \theta_t}.
	\end{align}
\end{theorem}

Theorem \ref{theorem:distrmu} shows, that the model averaging estimate is biased for    the true target parameter $\mu_{\text{true}}$, unless we have $\sum_{s=1}^r w_s \mu_s(\theta_s^*(\xi)) = \mu_{\text{true}}$. Hence we aim to minimize the asymptotic mean squared error of the model averaging estimate. Note, that the bias does not depend on the sample size, while the variance is of order $O(1/n)$.

\subsection{Optimal designs for  model averaging of non-nested models} \label{sec4}

\cite{alhorn_optimal_2019}  determined optimal designs for model averaging  minimizing the asymptotic mean squared error  of the estimate
calculated in a class of nested models under  local alternatives   and demonstrated
that  optimal designs lead to substantially more precise  model averaging estimates than commonly used designs in dose finding studies.
With the results of Section \ref{sec31} we can develop a more general concept of design of experiments
for model averaging estimation, which is applicable for non-nested models and in situations, where
the ``true'' model is not  contained in the set of candidate models used for model averaging.

To be precise, we consider the criterion
\begin{align}\label{asymp_mav_mse}
\Phi_{\text{mav}}(\xi, g) =
 \tfrac{1}{n} \sigma_w^2(\theta^*(\xi)) + \Big ( \sum_{s=1}^r w_s \mu_s(\theta_s^*(\xi)) - \mu_{\text{true}} \Big )^2 \approx \text{MSE} (\hat\mu_{\text{mav}}) ,
\end{align}
where $\mu_{\text{true}}$ is the  target parameter in the ``true''  model with  density $g$ and $\sigma_w^2(\theta^*(\xi))$  and  $ \theta_s^*(\xi) $
are defined  in \eqref{sigma_w} and \eqref{kl_min}, respectively.  Note that this criterion depends on the   ``true''  distribution  via $\mu_{\text{true}}$
and the best approximating parameters $\theta_{s}^*(\xi) = \theta_{s,g}^*(\xi) $.

For estimating the target parameter $\mu$ via a model averaging estimate of the form \eqref{mu_mav} most precisely a ``good" design $\xi$ yields small values of
the criterion function  $\Phi_{\text{mav}}(\xi, g)$. Therefore, for a given  finite set of candidate models $f_1,\ldots, f_r$  and weights $w_s, s=1,\ldots,r,$
 a design $\xi^*$ is called {\it  locally  optimal design for model averaging estimation of the parameter $\mu$}, if  it minimizes the function $\Phi_{\text{mav}}(\xi, g)$
 in \eqref{asymp_mav_mse}  in the class of  all approximate designs  on $\mathcal{X}$.
Here the term ``locally'' refers to the seminal paper of  \cite{chernoff_locally_1953} on optimal designs for nonlinear regression models,
because the optimality criterion still depends the unkown density $g( y \mid x) $.

A general approach to address this uncertainty  problem is a Bayesian approach based on a class of models for the density $g$.
To be precise, let   ${\cal G}$ denote a finite  set of potential  densities and let $\pi$ denote a probability distribution   on ${\cal G}$, then we
call  a design
\emph{Bayesian optimal design for model averaging estimation of the parameter} $\mu$ if it minimizes  the function
\begin{align} \label{bayes_mse}
\Phi_{\text{mav}}^\pi (\xi ) = \int_{\cal G} \Phi_{\text{mav}} (\xi,g )  d \pi (g) ~.
\end{align}
In  general, the set  ${\cal G} $  can  be constructed independently of the set of candidate models.  However, if there is not much prior information
available  one can construct a class of potential models ${\cal G}$ from the
candidate set as follows.
We denote the candidate set of   models in \eqref{candmod} by   $\mathcal{S}$. Each of these models depends on a unknown parameter $\theta_s$
and we denote by ${\cal F}_{f_s} \subset \Theta_{s}$ a  set  of possible parameter values for the model $f_{s}$. Now let
$\pi_{2}$ denote a prior distribution on ${\cal S}$ and for each $f_{s} \in {\cal S} $ let $\pi_{1} ( \cdot  \mid f_{s} ) $ denote a prior  distribution on ${\cal F}_{f_s}$.
Finally, we define
$
\mathcal{G} = \{(g,\theta ): g \in \mathcal{S}, \theta \in \mathcal{F}_{g} \}
$
 and  a prior
\begin{equation} \label{prior}
d \pi   (g  , \theta )=
d\pi_{1} ( \theta  \mid g )   ~d\pi_{2} (g)\;,
\end{equation}
then the criterion \eqref{bayes_mse} can be rewritten as
\begin{align} \label{bayes_msealt}
\Phi_{\text{mav}}^\pi (\xi ) = \int_{\cal S} \int_{{\cal F}_{g}} \Phi_{\text{mav}} (\xi,g) d\pi_1(\theta  \mid g) ~d \pi_2(g),
\end{align}
In  the finite sample study of the following section the set  ${\cal S} $ and  the set   $ \mathcal{F}_{g}$  (for any  $ g \in {\cal S} $) are finite, which  results in a finite set ${\cal G}$.

{Locally and Bayesian optimal designs for model averaging estimation have to be calculated numerically in all cases of practical interest. We will state now a necessary condition for the 
optimality of a given design with respect to the criterion   $\phi_{\text{mav}}^{\pi}$.  Note, that this criterion is not convex and therefore a  
sufficient condition cannot be derived.   In the following discussion we denote by
$A_s^{*} = A_s(\theta_{s,g}^*(\xi^{*}),\xi^{*})$   and $B_{st}^{*} = B_{st}(\theta_{s,g}^*(\xi^{*}),\theta_{t,g}^*(\xi^{*}),\xi^{*})$ 
the matrices  defined  in \eqref{As}  and \eqref{Bst}, respectively,  evaluated in $\xi^*$   and $\theta_{s,g} (\xi^{*})$. }

\begin{theorem}\label{bayes_mav_necess_cond}
If a  design $\xi^*$ is  Bayesian optimal for model averaging  estimation of the parameter $\mu$ with respect to the prior $\pi$, then
\begin{eqnarray} \label{nec_cond_bayes}
&&d_\pi  (x , \xi^*)   = \\
 && \int_{\mathcal{G}} \tfrac{1}{n} \sigma_g'(\xi^*,x) + 2 \Big ( \sum_{s=1}^r w_s \mu_s(\theta_{s,g}^*(\xi^*)) - \mu_{\text{true}} \Big) \sum_{s=1}^r w_s \Big ( \tfrac{\partial \mu_s(\theta_{s,g}^*(\xi^*))}{\partial \theta_s} \Big)^\top \theta_{s,g}'(\xi^*,x) d\pi(g)  \leq 0 \nonumber 
\end{eqnarray}
holds for all $x \in \mathcal{X}$, where the derivatives $\theta_s'(\xi^*,x)$ and $\sigma_g'(\xi^*,x)$ are given by
\begin{small}
\begin{align} \label{deriv_theta_star} \notag
	\theta_{s,g}'(\xi^*,x) &= - \Big (  \int \int g(y\mid t) \tfrac{\partial^2}{\partial \theta_s \partial \theta_s^\top} \log f_s (y\mid t,\theta_{s,g}^*(\xi^*)) dy d\xi^*(t) \Big )^{-1} \cdot \\ 
	& ~~~~~~~~~~~~~~~~~~~~~~~~~~~~~~
	\int g(y\mid x) \tfrac{\partial}{\partial \theta_s} \log f_s (y\mid x,\theta_{s,g}^*(\xi^*)) dy  \\
 \label{deriv_sigma}  
\sigma_g'(\xi^*,x) & = \sum_{s,t} w_s w_t \cdot \Big [ \Big ( \tfrac{\partial^2 \mu_s(\theta_{s,g}^*(\xi^*))}{\partial \theta_s \partial \theta_s^\top} \theta_{s,g}'(\xi^*,x) \Big)^\top ( A_s^*)^{-1} B_{st}^* 
(A_t^*)^{-1} \tfrac{\partial \mu_t(\theta_{t,g}^*(\xi^*))}{\partial \theta_t}   \\ \notag
& - \Big ( \tfrac{\partial \mu_s(\theta_{s,g}^*(\xi^*))}{\partial \theta_s} \Big )^\top \big ( (A_s^*)^{-1} h_{s,g}'(\xi^*,x) (A_s^*)^{-1} \big ) B_{st}^*
( A_t^*)^{-1} \tfrac{\partial \mu_t(\theta_{t,g}^*(\xi^*))}{\partial \theta_t}  \\ \notag
& + \Big ( \tfrac{\partial \mu_s(\theta_{s,g}^*(\xi^*))}{\partial \theta_s} \Big)^\top  (A_s^*)^{-1} h_{st,g}'(\xi^*,x) (A_t^*)^{-1} \tfrac{\partial \mu_t(\theta_{t,g}^*(\xi^*))}{\partial \theta_t}  \\ \notag
&- \Big ( \tfrac{\partial \mu_s(\theta_{s,g}^*(\xi^*))}{\partial \theta_s} \Big )^\top (A_s^*)^{-1}B_{st}^*
 \left(  (A_t^*)^{-1} h_{t,g}'(\xi^*,x) (A_t^*)^{-1} \right) \tfrac{\partial \mu_t(\theta_{t,g}^*(\xi^*))}{\partial \theta_t}  \\ 
& + \Big ( \tfrac{\partial \mu_s(\theta_{s,g}^*(\xi^*))}{\partial \theta_s} \Big )^\top (A_s^*)^{-1} B_{st}^*(A_t^*)^{-1} \tfrac{\partial^2 \mu_t(\theta_{t,g}^*(\xi^*))}{\partial \theta_t \partial \theta_t^\top} \theta_{t,g}'(\xi^*,x) \Big ], \notag
\end{align}
\end{small}
respectively. Here   the matrices $h_{st,g}'(\xi^*,x)$ and $h_{s,g}'(\xi^*,x)$  are given by 
\begin{small} 
\begin{align}\label{deriv_Bst} \notag
h_{st,g}'(\xi^*,x)  & = \int \int g(y\mid u) \left[ \tfrac{\partial^2 \log f_s(y \mid u,\theta_{s,g}^*(\xi^*))}{\partial \theta_s \partial \theta_s^\top} \theta_{s,g}'(\xi^*,x) \left( \tfrac{\partial \log f_t(y\mid u,\theta_{t,g}^*(\xi^*))}{\partial \theta_t} \right)^\top \right. \\ \notag
&\hspace{3cm} \left. + \tfrac{\partial \log f_s(y \mid u,\theta_{s,g}^*(\xi^*))}{\partial \theta_s} \left( \theta_{t,g}'(\xi^*,x)\right)^\top \left( \tfrac{\partial^2 \log f_t(y\mid u,\theta_{t,g}^*(\xi^*))}{\partial \theta_t \partial \theta_t^\top} \right)^\top \right] dy d\xi^*(u) \\ 
&\hspace{3cm} + B_{st}(\theta_{s,g}^*(\xi^*),\theta_{t,g}^*(\xi^*),\xi_x) - B_{st}(\theta_{s,g}^*(\xi^*),\theta_{t,g}^*(\xi^*),\xi^*) \\
 \label{deriv_As} \notag
h_s'(\xi^*,x) & = \int \int g(y \mid u) D_s(\theta_{s,g}^*(\xi^*)) (I_{p_s} \otimes \theta_{s,g}'(\xi^*,x)) dy d\xi^*(u) \\ 
&\hspace{3cm} + A_s(\theta_{s,g}^*(\xi^*),\xi_x) - A_s(\theta_{s,g}^*(\xi^*),\xi^*) 
\end{align}
\end{small}
where the matrix
\begin{align} \nonumber
D_s(\theta_s) = \left( 
\begin{array}{cccccccc}
\tfrac{\partial^3 \log f_s(y \mid x , \theta_s)}{\partial \theta_{s,1} \partial\theta_{s,1} \partial \theta_{s,1}} & \cdots & \tfrac{\partial^3 \log f_s(y \mid x , \theta_s)}{\partial \theta_{s,1} \partial \theta_{s,1} \partial \theta_{s,p_s}} &
 \cdots & \tfrac{\partial^3 \log f_s(y \mid x , \theta_s)}{\partial \theta_{s,1} \partial \theta_{s,p_s} \partial \theta_{s,1}} & \cdots & \tfrac{\partial^3 \log f_s(y \mid x , \theta_s)}{\partial \theta_{s,1} \partial \theta_{s,p_s} \partial\theta_{s,p_s}} \\
\tfrac{\partial^3 \log f_s(y \mid x , \theta_s)}{\partial \theta_{s,2} \partial \theta_{s,1} \partial \theta_{s,1}} & \cdots & \tfrac{\partial^3 \log f_s(y \mid x , \theta_s)}{\partial \theta_{s,2}\partial\theta_{s,1}\partial\theta_{s,p_s}} 
& \cdots & \tfrac{\partial^3 \log f_s(y \mid x , \theta_s)}{\partial \theta_{s,2} \partial\theta_{s,p_s} \partial\theta_{s,1}} & \cdots & \tfrac{\partial^3 \log f_s(y \mid x , \theta_s)}{\partial \theta_{s,2} \partial \theta_{s,p_s} \partial \theta_{s,p_s}} \\
\vdots &  & \vdots 
& & \vdots & & \vdots \\
\tfrac{\partial^3 \log f_s(y \mid x , \theta_s)}{\partial \theta_{s,p_s}\partial\theta_{s,1}\partial\theta_{s,1}} & \cdots & \tfrac{\partial^3 \log f_s(y \mid x , \theta_s)}{\partial \theta_{s,p_s}\partial\theta_{s,1}\partial\theta_{s,p_s}} 
& \cdots & \tfrac{\partial^3 \log f_s(y \mid x , \theta_s)}{\partial \theta_{s,p_s}\partial\theta_{s,p_s}\partial \theta_{s,1}} & \cdots & \tfrac{\partial^3 \log f_s(y \mid x , \theta_s)}{\partial \theta_{s,p_s}\partial\theta_{s,p_s}\partial\theta_{s,p_s}}
\end{array}
\right)
\end{align}
contains the third derivatives of the log-likelihood with respect to  the parameters $\theta_s = (\theta_{s,1},\ldots,\theta_{s,p_s})^\top$. 
Moreover, there is equality in \eqref{nec_cond_bayes} for all support points of the optimal design.
\end{theorem}

\begin{figure}[t]
\centering
\includegraphics[width=0.4\textwidth]{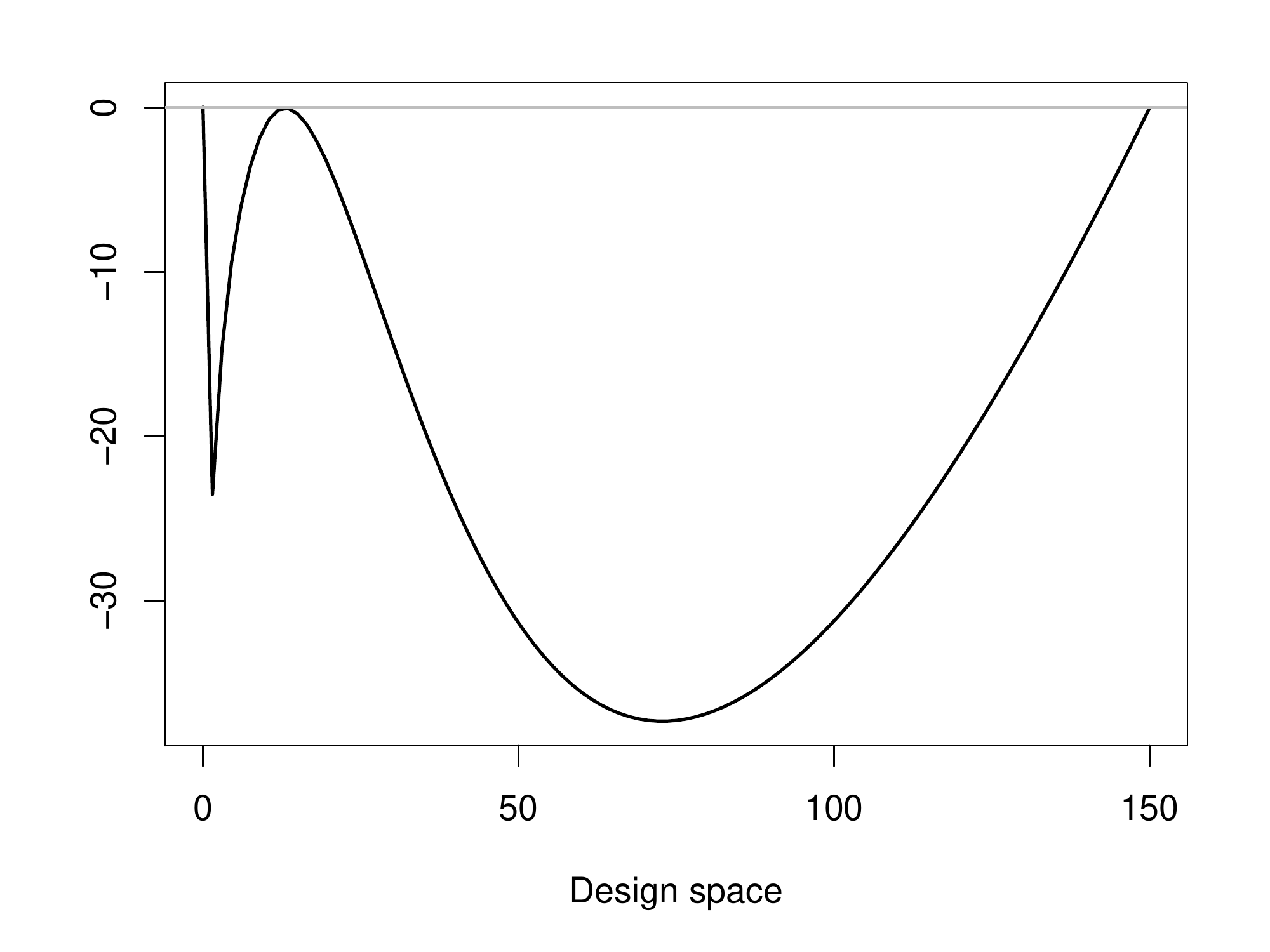}
\caption{\it Necessary condition of Theorem \ref{bayes_mav_necess_cond} for the optimal design \eqref{optdesillu}.}
\label{fig_deriv_ill}
\end{figure}

\begin{example} {\rm 
We illustrate the application of Theorem \ref{bayes_mav_necess_cond} 
for  
 regression models  of the from  \eqref{regression_model}
with  centred   normal distributed errors.    As regression functions we use the log-linear and the Emax model and their parameter specifications given in 
Table \ref{tab1}. Then, the locally optimal design for estimation of the $\text{ED}_{0.4}$ in the log-linear model $f_1$  
and in  the Emax model $f_2$  
are  given by
\begin{align} \label{vgldes2}
	\xi_2 & = \left\{
	0 , 4.051 , 150 ;~
	0.339 , 0.5 , 0.161	
	 \right\},
\end{align}
and 
$\left\{
	0 , 18.75 , 150 ;~
	0.25 , 0.5 , 0.25	
	 \right\}, $
 respectively [see  \cite{dette_optimal_2010}].
For sample  size  $n=100$ we  determine a Bayesian optimal design for model averaging estimation of the $\text{ED}_{0.4}$  (with uniform weights)  
 with respect to the criterion \eqref{bayes_mse}. The set of possible models is given by $\mathcal{G} = \{f_1,f_2\}$ with parameters specified in Table \ref{tab1}, and we   choose a uniform prior on this set. The optimal design has been calculated numerically using the COBYLA algorithm (see \cite{powell_direct_1994}) 
 and   is given by
\begin{align} \label{optdesillu}
\xi_{12}^* = \left\{
	0 , 13.026 , 150 ;~
	0.281, 0.498, 0.220	
	 \right\}.
\end{align}
The necessary condition of Theorem \ref{bayes_mav_necess_cond} is satisfied as illustrated in Figure \ref{fig_deriv_ill}). 
 Note that the  design $\xi_{12}^*$  can be considered as  a compromise between the locally optimal designs 
for the individual models and that $\xi_{12}^*$  would not be optimal if the inequality was not satisfied.   
}
\end{example}
We conclude noting that the optimality criteria proposed in this section have been derived for  model averaging  estimates with fixed  weights.
The asymptotic  theory  presented here cannot be easily adapted to estimates using data-dependent (random) weights  (as considered in  Section \ref{sec2}), because
it is difficult to get an explicit  expression for the asymptotic distribution, which is not normal in general.
Nevertheless, we will demonstrate in the following section that designs minimizing the  mean squared error  of   model averaging  estimates with fixed  weights
will also yield a substantial improvement in model averaging estimation with smooth AIC-weights and in  estimation after  model selection.

\section{Bayesian optimal designs for model averaging  } \label{sec5}
\def\theequation{4.\arabic{equation}}
\setcounter{equation}{0}

We will demonstrate  by means of a simulation study that the performance of all considered estimates can be 
 improved  substantially  by the choice of an appropriate design. For this purpose  
 we consider   the same  situation  as  in Section \ref{sec2}, that is  
 regression models  of the from  \eqref{regression_model}
with  centred   normal distributed errors.  We also consider the two different candidate sets
${\cal S}_{1}$ and  ${\cal S}_{2}$
defined in \eqref{set1} (log-linear, Emax and quadratic model) and \eqref{set2} (log-linear, Emax and exponential model), respectively.

Using the criterion introduced in  Section \ref{sec3} we now determine a Bayesian optimal design for model averaging estimation of the $\text{ED}_{0.4}$
with uniform weights from  $n=100$ observations. We require a prior distribution for the unknown density $g$, and we use  a distribution of the  form
\eqref{prior} for this purpose. To be precise, let $f_{s}(y   \mid x, \theta_{s})$ denote the density of a normal distribution with mean $\eta_{s} (x, \vartheta_{s})$
and variance $\sigma^{2}_{s} =0.1$ ($s=1, \ldots , r$), where  the mean functions are given in Table \ref{tab1}.
 As the criterion \eqref{bayes_msealt} does  not depend on the intercept  $\vartheta_{s1}$, these are not varied and taken from Table \ref{tab1}.
  For  each of  the other  parameters  we use three different  values: the values specified in Table \ref{tab1}
 and a $10\%$ larger and smaller value of this parameter. 
\begin{align} \label{candidate_set}
	{\cal F}_{f_{1}} &= \{ (0, \vartheta_{12}, \vartheta_{13}) :  \vartheta_{12} = 0.0797  \pm 10\% , \vartheta_{13} =  1  \pm 10\% \},  \\
	{\cal F}_{f_{2}} &= \{(0,\vartheta_{22}, \vartheta_{23}) : \vartheta_{22} = 0.467 \pm 10\%, \vartheta_{23} = 25 \pm 10\% \} ,\notag \\
{\cal F}_{f_{3}}  &= \{(-0.08265, \vartheta_{32}, \vartheta_{33}) : \vartheta_{32} = 0.08265 \pm 10 \%, \vartheta_{33} = 85 \pm 10\% \} , \notag \\
	{\cal F}_{f_{4}}  &= \{(0, \vartheta_{42}, \vartheta_{43}) : \vartheta_{42} = 0.00533 \pm 10\%, \vartheta_{43} = -0.00002 \pm 10\% \}.
	 \notag
\end{align}

\subsection{Models of similar shape} \label{sec52}
We will first consider  the candidate set ${\cal S}_{1} =\{ f_{1},f_{2},f_{4} \} $
consisting of the log-linear, the Emax and the quadratic model.  For the definition of the prior distribution \eqref{prior}
in the criterion  \eqref{bayes_msealt}  we consider a uniform distribution
$\pi_{2} $ on the set ${\cal S}_{1} $ and a uniform prior  $\pi_{1} ( \cdot  \mid f_{s} ) $ on each set  ${\cal F}_{f_{s}} $  in \eqref{candidate_set}
($s=1,2,4$).
The Bayesian optimal design for model averaging estimation of the $\text{ED}_{0.4}$  minimizing the criterion \eqref{bayes_msealt}
  is given by 
\begin{align} \label{optquad}
\xi_{{\cal S}_1}^* =
\left\{
	0 ,  18.310,  67.102, 150;~
	0.205, 0.290, 0.281, 0.224
\right\}~.
\end{align}
We will compare this design  with the
 design
\begin{align} \label{vgldes1}
	\xi_1 = \left\{
	0 ,  10 , 25 , 50 , 100 , 150 ;~
	1/6 ,  1/6 ,  1/6 , 1/6 , 1/6 , 1/6	
 \right\},
\end{align}
proposed  in \cite{pinheiro_design_2006} for a  a similar setting (this design has also been used in  Section \ref{sec2})
and the locally optimal design for the  estimation of the $\text{ED}_{0.4}$ in the log-linear model  given by \eqref{vgldes2}.
Results for the  locally optimal designs for  estimation of the $\text{ED}_{0.4}$ in the Emax and  exponential model
are similar and omitted for the sake of brevity.
We use the same   setup as in Section \ref{sec2}.    Only results for the sample size 
$n=100$ are reported   and  results for other sample sizes are available from the authors.

 \begin{table}[t]
 \centering
   \renewcommand{\arraystretch}{.9}
 \small
 \begin{tabular}{| c | c || c|c|c|}
 \hline
model & design  & uniform  weights &  smooth AIC-weights & model selection \\
 \hline
  &  \eqref{optquad} & \textbf{177.472} & \textbf{165.981} & \textbf{173.548} \\ 
$f_1$   & \eqref{vgldes1}  & 223.291 & 218.990 & 285.062 \\
  & \eqref{vgldes2}  & 185.251 & 184.77 & 340.698 \\  \hline
 &  \eqref{optquad} & \textbf{142.085} & \textbf{153.745} & \textbf{170.059} \\ 
  $f_2$   & \eqref{vgldes1}   & 189.785 & 203.796 & 251.836 \\
  & \eqref{vgldes2}  & 501.814 & 501.394 & 1162.654 \\ \hline
 &  \eqref{optquad} &  \textbf{160.039} & \textbf{195.116} & \textbf{299.365} \\ 
 $f_4$   & \eqref{vgldes1}  &  190.662 & 244.558 & 391.443 \\
  & \eqref{vgldes2}  &  404.716 & 427.548 & 1396.051 \\
  \hline
    \hline
 &  \eqref{optquad}  &\textbf{1058.655} & \textbf{766.408} & \textbf{606.752} \\ 
   $f_3$    & \eqref{vgldes1}    & 1109.622 & 856.484 & 729.912 \\
  & \eqref{vgldes2} & 3184.11 & 3413.566 & 4102.964 \\
   \hline
 \end{tabular}
   \renewcommand{\arraystretch}{1}
 \caption{\it  Simulated    mean squared errors of different estimates of the $\text{ED}_{0.4}$ for different experimental designs. The set of candidate models is   $\mathcal{S}_1 = \{f_1, f_2, f_4\}$. Left column: model averaging estimate with uniform weights; middle column: model averaging estimate with smooth AIC-weights; right column: estimate  after  model selection.
}
\label{tab4}
 \end{table}

The corresponding results  are given in Table \ref{tab4}, where we  use  the models $f_{1}, f_{2},f_{3}$ and $f_{4}$  from Table \ref{tab1}
to generate the data (note that the model $f_{3}$  is not in the candidate set used for model averaging and model selection). The different columns
represent the different estimation methods  (left column: model averaging with uniform  weights; middle column:  smooth AIC-weights, right column: model selection).
The numbers printed in boldface indicate the minimal  mean squared error for each estimation method  obtained from the different experimental designs.
First, we consider the situation, where the data generating  model is contained in the set of candidate models ${\cal S}_{1}= \{ f_{1}, f_{2},f_{4}\}$
corresponding to the  upper part of the table. We observe that in this case model averaging yields better results than estimation after  model selection and
this superiority is independent of the design under consideration.
 Compared to the designs $\xi_1$ and $\xi_2$ the Bayesian optimal design $\xi_{{\cal S}_1}^*$  for model averaging with uniform  weights  improves the efficiency
of  all estimation techniques. For example, when data is generated using the log-linear model $f_1$  the
mean squared error of the model averaging estimate with   uniform weights  is reduced  by $20.5\%$ and $4.2\%$, when the optimal design is
used instead of  the designs $\xi_1$ or $\xi_2$,  respectively. This improvement is remarkable as the design $\xi_2$ is locally optimal for estimating
the ED$_{0.4}$ in the model $f_{1}$ and data  is generated from this model.
 In other cases  the improvement is even more visible. For example, if data is generated by the model $f_{2}$ the improvement
 in model averaging estimation with uniform  weights is
 $25.1\%$ and $71.7\%$ compared to the designs $\xi_{1}$ and $\xi_{2}$ defined in \eqref{vgldes1} and   \eqref{vgldes2}.   Moreover, although the designs are  constructed for model averaging with
 uniform weights they also yield substantially more accurate
 model averaging estimates  with  smooth AIC-weights  and  a more precise estimate after model selection.
 For example, if the data is generated from model $f_{1}$  the mean squared
 error  is reduced by $24.2\%$   and  by  $10.2\%$ for estimation with smooth AIC-weights
 and  by $39.1\%$ and $49.1\%$   for estimation after  model selection, respectively. Similar  results can be observed for the models
 $f_{2}$ and $f_{4}$.

Next, we consider the case  where the data  is generated from  the exponential model $f_{3}$, which is not contained in the candidate set ${\cal S}_{1}$.
The efficiency of  all three estimates  improves substantially by the use of the Bayesian optimal
 design $\xi_{{\cal S}_1}^*$. Interestingly, the improvement is less pronounced  for model averaging with uniform  weights ($4.6\%$ and $66.8\%$
 compared to the designs $\xi_{1}$ and $\xi_{2}$ in  in \eqref{vgldes1} and   \eqref{vgldes2}, respectively) than for smooth AIC-weights ($10.5\%$ and $77.5\%$)  and
 estimation after  model selection  ($16.9\%$ and $85.2\%$).

 Summarizing, our numerical results show that the Bayesian optimal design for model averaging estimation of the $\text{ED}_{0.4}$ yields  a substantial
 improvement of  the mean   squared error of the model averaging estimate with  uniform
 weights ($4.2\%$-$71.7\%$),
smooth AIC-weights  ($10.2\%$-$77.5\%$) and the estimate after  model selection ($16.9\%$-$85.4\%$) for all four models under consideration.

\subsection{Models of different shape} \label{sec53}

We will now consider the second candidate set ${\cal S}_{2}$ consisting of the log-linear  ($f_{1} $)
 the Emax ($f_{2}$) and the exponential model  ($f_{3}$).
For the definition of the prior distribution \eqref{prior}
in the criterion  \eqref{bayes_msealt}  we use a uniform distribution
  $\pi_{2} $ on the set ${\cal S}_{2} $ and a uniform prior  $\pi_{1} ( \cdot  \mid f_{s} ) $ on each set  ${\cal F}_{f_{s}} $ ($s=1,2,3$) in \eqref{candidate_set}. For this choice
    the Bayesian optimal design for model averaging estimation of the $\text{ED}_{0.4}$ is given by
\begin{align} \label{optdes1}
\xi^* _{{\cal S}_2}= \left\{
0 , 10.025,  77.746,  84.556, 150 ; ~
0.192, 0.212, 0.198, 0.189, 0.208
\right\},
\end{align}
and   has (in comparison to the design $\xi_{{\cal S}_1}^*$ in Section \ref{sec52}) five instead of four support points.

   \begin{table}[t]
 \centering
   \renewcommand{\arraystretch}{.9}
 \small
 \begin{tabular}{| c | c || c|c|c|}
 \hline
 \multicolumn{2}{|c||}{} & \multicolumn{3}{|c|}{estimation method} \\
\cline{3-5}
model & design  & uniform  weights &  smooth AIC-weights & model selection \\
 \hline
  &  \eqref{optdes1} &  \textbf{654.914} & \textbf{279.257} & \textbf{274.016} \\ 
$f_1$  & \eqref{vgldes1}   & 712.404 & 340.254 & 353.707 \\
  &\eqref{vgldes2}  & 770.705 & 410.715 & 413.676 \\ \hline
  &\eqref{optdes1} & 546.105 & 283.757 & \textbf{250.719} \\ 
  $f_2$  & \eqref{vgldes1}   & \textbf{517.963}  & \textbf{267.967}   & 286.272 \\
&\eqref{vgldes2} & 1098.323 & 962.257 & 1701.569 \\  \hline
  &\eqref{optdes1} & 910.372 & \textbf{742.507} & \textbf{699.612} \\ 
  $f_3$  &  \eqref{vgldes1}  &  \textbf{871.362} & 766.140 & 802.763 \\
  & \eqref{vgldes2} &  1505.693 & 1774.895 & 2592.261 \\ 
  \hline  \hline
 &  \eqref{optdes1}  &  \textbf{159.899} & \textbf{278.409} & \textbf{347.187} \\ 
    $f_4$  &  \eqref{vgldes1}    &  208.628 & 298.315 & 419.651 \\
  & \eqref{vgldes2} &  522.652 & 610.198 & 1907.066 \\ 
  \hline
 \end{tabular}
   \renewcommand{\arraystretch}{1}
 \caption{\it   Simulated    mean squared errors of different estimates of the $\text{ED}_{0.4}$ for different experimental designs. The set of candidate models is   $\mathcal{S}_2 = \{f_1, f_2, f_3\}$. Left column: model averaging estimate with uniform weights; middle column: model averaging estimate with smooth AIC-weights; right column: estimate after  model selection.}
\label{tab5}
 \end{table}

 The  simulated mean squared errors of the three estimates under different designs
 are given in Table \ref{tab5}. We observe  again that compared to the designs $\xi_{1}$ and $\xi_{2} $   in \eqref{vgldes1} and   \eqref{vgldes2} the  Bayesian optimal design
  $\xi^* _{{\cal S}_2}$
 improves most  estimation techniques substantially.  However, if  model averaging with uniform weights is used and data  is
 generated by model {$f_{2}$ or $f_3$}, the mean squared  error  of the model averaging estimate from the optimal design is   {$5.4\%$ and $4.5\%$}
 larger than the mean squared error obtained by the design $\xi_{1}$, respectively. For model averaging with smooth AIC-weights {and data being generated from model $f_2$} this difference
 is  {$5.9\%$}. Overall, the  reported results   demonstrate  a substantial improvement in efficiency
  by usage  of the Bayesian optimal design independently of the estimation method.
If the Bayesian optimal design is used, estimation after  model selection  yields the smallest mean squared error if the data is generated from a
 model of the candidate set ${\cal S}_{2}$. On the other hand, if data is generated from model $f_{4} \not \in {\cal S}_{2}$ model averaging with equal
 weights shows the best performance.

\medskip

 Summarizing, our numerical results show that compared to the designs $\xi_1$ and $\xi_2$  the design $\xi^* _{{\cal S}_2}$ reduces the mean squared error of  model averaging estimates with uniform weights up to  $69.4\%$. Furthermore,  for  smooth AIC-weights and estimation
 after  model selection  the  reduction can be even larger and is up to $70.5\%$ and $85.3\%$, respectively.
  These improvements hold also for the quadratic model  $f_{4}$, which is not contained in  the candidate set ${\cal S}_{2}$ used in the definition
  of the optimality criterion.

\section{Conclusions} \label{sec7}

In this paper we derived the asymptotic distribution of the  frequentist model averaging estimate with fixed weights  from a class of not necessarily nested models.
We neither assume that this class contains the ``true'' model. We use these results to determine Bayesian optimal designs for model averaging,
which can improve the estimation accuracy  of the estimate substantially. Although these designs are constructed for model averaging with fixed weights,  they also
yield a substantial improvement of accuracy for model averaging with data dependent weights and for estimation after model selection.

We also demonstrate that   the superiority of model averaging against estimation after model selection depends sensitively on the class of competing models, which is used in the
model averaging procedure. If the competing models are similar (which means that a given model from the class can be well approximated by all other models), then model averaging should
be preferred. Otherwise, we observe advantages for estimation after model selection, in particular, if the signal to noise ratio is small.

Although, the new designs show a very good performance for estimation after model selection and for model averaging with data dependent weights, it is of interest to develop optimal
designs, which address the specific issues of data dependent weights in the estimates.  This is a very challenging problem for future research
as there is no simple expression of the asymptotic mean squared error of these estimates. A first approach to solve this problem is an adaptive one and a further interesting and very challenging question of future research is to improve the accuracy of adaptive designs.



\spacingset{1.5}

\section{Technical assumptions and proofs}  \label{app}
\def\theequation{A.\arabic{equation}}
\setcounter{equation}{0}

\paragraph{6.1 Assumptions} \label{app1}
Following \cite{white_1982} we assume:
\begin{itemize}
	\item[(A1)] The random variables $Y_{ij}, i=1,\ldots,k, j=1,\ldots,n_i$  are independent. Furthermore,
	$Y_{i1},\ldots ,Y_{in_{i}}$	have a common distribution function
	with a measurable  density $g(\cdot  \mid x_{i})$ with respect to a dominating measure $\nu$.
	\item[(A2)]  The distribution  function of
	each  candidate model $s\in\{1,\ldots,r\}$   has a  measurable  density $f_s(\cdot  \mid x,\theta_s)$ with respect to $\nu$ (for all $\theta_s \in \Theta_s$)
	 that is continuous in $\theta_s$.
	\item[(A3)] For all $x \in \mathcal{X}$ the expectation
	 $\E(\log(g(Y\mid x)))$ exists (where expectation is taken with respect to $g(\; \cdot  \mid x) $) and for each candidate model
	 the function $y \mapsto |\log f_s(y \mid x,\theta_s)|$ is dominated by a function that is integrable with respect to  $g(\; \cdot  \mid x) $  and does not depend on $\theta_s$.
	 Furthermore the Kullback-Leibler divergence \eqref{kldist} has a unique minimum $\theta_{s,g}^*(\xi)$ defined in \eqref{kl_min}
	and $\theta_{s,g}^*(\xi)$ is an interior point of $\Theta_s$.
	\item[(A4)] For all $x \in \mathcal{X}$ the function $ y \mapsto \tfrac{\partial \log f_s(y \mid x,\theta_s)}{\partial \theta_{s}}$ is a measurable function for all $\theta_s \in \Theta_s$ and   continuously differentiable with respect to $\theta_s$ for all $y\in \mathbb{R}$.
	\item[(A5)]  The entries of the (matrix valued) functions
	$
	\tfrac{\partial^2 \log f_s(y \mid x,\theta_s)}{\partial \theta_s \partial \theta_s^{\top}}
	$, $
	\tfrac{\partial \log f_s(y \mid x,\theta_s)}{\partial \theta_s} \big ( \tfrac{\partial \log f_t(y \mid x,\theta_t)}{\partial \theta_t} \big)^{\top}
	$
	are dominated by integrable functions with respect to $g(\; \cdot \mid x)$ for all $x\in\mathcal{X}$ and $\theta_s \in \Theta_s$.
	\item[(A6)] The matrices  $B_{ss}(\theta_s^*(\xi),\theta_s^*(\xi), \xi )$ and $A_s(\theta_s^*(\xi), \xi )$ in  \eqref{As} and  \eqref{Bst}  are  nonsingular.
	\item[(A7)] The functions $ \theta_s \mapsto  \mu_s ( \theta_s) $ are once continuously differentiable.
\end{itemize}

\paragraph{6.2  Proof of Theorem \ref{theorem:distrmu}.}

By equation (A.2) in \cite{white_1982} we have
\begin{align} \label{A.2}
	\sqrt{n} (\hat{\theta}_{n,s} - \theta_s^*(\xi)) + A_s^{-1}(\theta_s^*(\xi)) \tfrac{1}{\sqrt{n}} \sum_{i=1}^k \sum_{j=1}^{n_i} \tfrac{\partial \log f_s(Y_{ij}  \mid x_i,\theta_s^*(\xi))}{\partial \theta_s} \overset{p}{\longrightarrow} 0,
\end{align}
where $\overset{p}{\longrightarrow}$ denotes convergence in probability
(note that the matrix  $A_s(\theta_s^*)= A_s(\theta_s^*,\xi )$ is nonsingular by assumption).
An application of  the multivariate central limit theorem now leads to
\begin{align} \label{as_loglik}
	\tfrac{1}{\sqrt{n}} \left(
		\begin{array}{c}
		\sum_{i=1}^k \sum_{j=1}^{n_i} \tfrac{\partial \log f_1(Y_{ij}\mid x_i,\theta_1^*(\xi))}{\partial \theta_1} \\
		\vdots \\
		\sum_{i=1}^k \sum_{j=1}^{n_i} \tfrac{\partial \log f_r(Y_{ij}\mid x_i,\theta_r^*(\xi))}{\partial \theta_r}
		\end{array}
		\right) \overset{\mathcal{D}}{\longrightarrow} \mathcal{N}\left( 0, \left( \begin{array}{cccc}
		B_{11}  & \hdots & B_{1r}  \\
		 \vdots & \ddots & \vdots \\
		B_{r1}  &  \hdots & B_{rr}
		\end{array} \right) \right),
\end{align}
where  $B_{{st}} = B_{{st}} (\theta_s^*(\xi), \theta_t^*(\xi), \xi )$
is defined in \eqref{Bst}.
Combining \eqref{A.2} and \eqref{as_loglik} we obtain the weak convergence of the vector 	
	$\hat{\theta}_n = (\hat{\theta}_{n,1}^{\top},\ldots,\hat{\theta}_{n,r}^{\top})^{\top}$,
	that is
	$
		\sqrt{n} (\hat{\theta}_n - \theta^*(\xi)) \overset{\mathcal{D}}{\longrightarrow} \mathcal{N}(0,\Sigma),
	$
	where   $\Sigma = (\Sigma _{st})_{s,t=1, \ldots ,r}$ is a block matrix with 
	entries 	
	$
		\Sigma_{{st}} = A_s^{-1}(\theta_s^*(\xi)) B_{st} (\theta_s^*(\xi),\theta_t^*(\xi))  A_t^{-1}(\theta_t^*(\xi)) $
		 ($s,t=1, \ldots , r$)
{and the vector $\theta_s^*(\xi)$  is given by
$	\theta^*(\xi) = (\theta_1^*(\xi)^\top,\ldots,\theta_r^*(\xi)^\top)^\top$.

Next, we define for  the parameter vector
$\theta^{\top} = (\theta_1^{\top},...,\theta_r^{\top}) \in \mathbb{R}^{\sum_{s=1}^r p_s}	$
	 the projection $\pi_s$  by  $\pi_s\theta := \theta_s $ and the vector
	$\tilde{\mu}(\theta) = \big (
				\mu_1(\pi_1 \theta) , \ldots ,
		\mu_r(\pi_r \theta)
	\big )^{T}$
with  derivative
\begin{align} \label{deriv_mutilde}
	\mu_{\theta}' =
	\left(
	\begin{array}{ccccc}
		\left( \tfrac{\partial \mu_1(\theta_1)}{\partial \theta_1} \right)^{\top} & 0 & \hdots & & 0 \\
		0 & \left( \tfrac{\partial \mu_2(\theta_2)}{\partial \theta_2} \right)^{\top} & 0 & \hdots & 0 \\
		0 & \hdots & & 0 & \left( \tfrac{\partial \mu_r(\theta_r)}{\partial \theta_r} \right)^{\top}
	\end{array}
	\right).
\end{align}
An application of the Delta method   shows  that
	$
		\sqrt{n} (\tilde{\mu}(\hat{\theta}_{n}) - \tilde{\mu}(\theta^*(\xi))) \overset{\mathcal{D}}{\longrightarrow} \mathcal{N} \big (0, \mu_{\theta^*(\xi)}' \Sigma ( \mu_{\theta^*(\xi)}' )^{\top} \big ).
	$	
The assertion  finally  follows from the continuous mapping theorem observing the representation
$
\hat{\mu}_{\text{mav}} = \left( w_1, \hdots , w_r \right) \tilde{\mu}(\hat{\theta}_n).
$

\paragraph{6.3 Proof of Theorem \ref{bayes_mav_necess_cond}.}

{Throughout this proof we assume that integration and differentiation are interchangeable.  Following 
the arguments in \cite{pukelsheim_optimal_2006}, Chapter 11, a Bayesian optimal design $\xi^{*}$ for model averaging estimation of the parameter $\mu$ satisfies the inequality
\begin{equation}\label{nec_cond2}
-\int_{\cal{G} } D \Phi_{\text{mav}}(\xi^*,g)(\xi_x - \xi^*) d\pi(g) \leq 0
\end{equation}
for all $x \in \mathcal{X}$, where $D \Phi_{\text{mav}}(\xi^*,g)(\xi_x - \xi^*)$ denotes the directional derivative of the function $\Phi_{\text{mav}}$ evaluated in the optimal design $\xi^*$ in direction $\xi_x - \xi^*$ and $\xi_x$ denotes the Dirac measure at the point $x \in {\cal X}$. 
}
{To calculate  the derivative we  start with the derivative of the parameter $\theta_{s,g}^*(\xi)$ defined in \eqref{kl_min} and 
 define 
$\theta_{s,g}(\alpha) := \theta_{s,g}^*(\xi_{\alpha})$ for $\xi_{\alpha} = \alpha\xi_x + (1-\alpha) \xi^*$. Note that 
  $\theta_{s,g}(\alpha)$ is the solution of the equation
\begin{align} \label{thetas_solution}
F_{s,g}(\alpha,\theta_{s}) = - \int \int g(y\mid t) \tfrac{\partial}{\partial \theta_s} \log f_s (y\mid t,\theta_s) dy d\xi_{\alpha}(t) = 0,
\end{align}
and that the derivatives of the left hand side are given by 
\begin{align}
\tfrac{\partial F_{s,g}}{\partial \alpha} &= - \int \int g(y\mid t) \tfrac{\partial}{\partial \theta_s} \log f_s (y\mid t,\theta_s) dy d(\xi_x-\xi^*)(t), \\
\tfrac{\partial F_{s,g}}{\partial \theta_s}  &= - \int \int g(y\mid t) \tfrac{\partial^2}{\partial \theta_s  \partial  \theta_s^\top} \log f_s (y\mid t,\theta_s) dy d\xi_{\alpha}(t).
\end{align}
By the implicit function theorem we get
$
\tfrac{\partial \theta_{s,g}(\alpha)}{\partial \alpha}  = - \big ( \tfrac{\partial F_{s,g}}{\partial \theta_s} \big )^{-1} \tfrac{\partial F_{s,g}}{\partial \alpha}
$
and hence
\begin{align*} 
\left. \tfrac{\partial}{\partial \alpha} \theta_{s,g}(\alpha) \right\vert_{\alpha=0} &= - \Big ( \int \int g(y\mid t) \tfrac{\partial^2}{\partial \theta_s \partial \theta_s^\top} \log f_s (y\mid t,\theta_{s,g}^*(\xi^*)) dy d\xi^*(t) \Big )^{-1} \cdot \\ 
&\hspace{3ex} \int \int g(y\mid t) \tfrac{\partial}{\partial \theta_s} \log f_s (y\mid t,\theta_{s,g}^*(\xi^*)) dy d(\xi_x-\xi^*)(t) 
=\theta_s'(\xi^*,x),
\end{align*}
where $\theta_s'(\xi^*,x)$ is defined in \eqref{deriv_theta_star}. }
{Consider now the directional derivative of the matrix $B_{st}$ defined in \eqref{Bst}. An application of chain and product rule gives 
\begin{align} \nonumber
& \tfrac{\partial B_{st}(\theta_{s,g}^*(\xi_{\alpha}),\theta_{t,g}^*(\xi_{\alpha}),\xi_{\alpha})}{\partial \alpha} \Big \vert_{\alpha=0} \\ \notag
&= \Big ( \int \int g(y\mid u) \tfrac{\partial}{\partial \alpha} \Big ( \tfrac{\partial \log f_s(y\mid u,\theta_{s,g}^*(\xi_{\alpha}))}{\partial \theta_s} \Big ( \tfrac{\partial \log f_t(y\mid u,\theta_{t,g}^*(\xi_{\alpha}))}{\partial \theta_t} \Big )^\top \Big ) dy d\xi_{\alpha}(u)  \\ \notag
&  + \int \int g(y \mid u) \tfrac{\partial \log f_s(y\mid u,\theta_{s,g}^*(\xi_{\alpha}))}{\partial \theta_s} \Big ( \tfrac{\partial \log f_t(y\mid u,\theta_{t,g}^*(\xi_{\alpha}))}{\partial \theta_t} \Big )^\top dy d(\xi_x - \xi^*)(u) \Big ) \Big \vert_{\alpha=0}  \notag
=h_{st,g}'(\xi^*,x),
\end{align}
where $h_{st,g}'(\xi^*,x)$ is defined in \eqref{deriv_Bst}. In a similar way the derivative of the matrix $A_s$ defined in \eqref{As} can be determined.
 First, using the chain rule, we observe that
with $\theta_{s,g}'(\xi^*,x) = (\theta_{s,g,1}'(\xi^*,x),\cdots,\theta_{s,g,p_s}'(\xi^*,x))^\top$ 
\begin{align*}
\tfrac{\partial}{\partial \alpha} \left. \left( \tfrac{\partial^2 \log f_s(y \mid x,\theta_{s,g}^*(\xi_{\alpha}))}{\partial \theta_s \partial \theta_s^{T}} \right) \right\vert_{\alpha=0} 
&= D_s(\theta_{s,g}^*(\xi^*)) (I_{p_s} \otimes \theta_{s,g}'(\xi^*,x)) ,
\end{align*}
where $D_s$ is  defined in Theorem \ref{bayes_mav_necess_cond}. 
We now observe, that 
$
\tfrac{\partial A_s(\theta_{s,g}^*(\xi_{\alpha}),\xi_{\alpha})}{\partial \alpha} \big \vert_{\alpha=0} = 
 h_{s,g}'(\xi^*,x) $,  where $ h_{s,g}'(\xi^*,x)$ is defined in \eqref{deriv_As}.}
Noting, that 
\begin{align} \label{deriv_gradmu}
\left. \tfrac{\partial}{\partial \alpha} \tfrac{\partial \mu_s(\theta_{s,g}^*(\xi_{\alpha}))}{\partial \theta_s} \right\vert_{\alpha=0} = \tfrac{\partial^2 \mu_s(\theta_{s,g}^*(\xi^*))}{\partial \theta_s \partial \theta_s^\top} \theta_{s,g}'(\xi^*,x),
\end{align}
equation \eqref{deriv_sigma} results by  an application of the product rule and combination of the derivatives given above.
Finally, we have 
\begin{small}
\begin{align*}
 \tfrac{\partial }{\partial\alpha} \Big ( \sum_{s=1}^r w_s \mu_s(\theta_{s,g}^*(\xi)) - \mu_{\text{true}} \Big )^2 \Big \vert_{\alpha=0} = 2 \Big ( \sum_{s=1}^r w_s \mu_s(\theta_{s,g}^*(\xi^*)) - \mu_{\text{true}} \Big ) \sum_{s=1}^r w_s \Big ( \tfrac{\partial \mu_s(\theta_{s,g}^*(\xi^*))}{\partial \theta_s} \Big )^\top \theta_{s,g}'(\xi^*,x),
\end{align*}
\end{small}
and \eqref{nec_cond_bayes} follows.

The proof that there is  equality in \eqref{nec_cond_bayes} for all support points 
of the optimal design $\xi^{*} $ follows by a standard argument 
and the details are omitted for the sake of brevity.

\bigskip

{\bf Acknowledgements} This work has also been supported in part by the
Collaborative Research Center ``Statistical modeling of nonlinear
dynamic processes'' (SFB 823, Teilprojekt C2, T1) of the German Research Foundation
(DFG).

\setstretch{1.15}
\setlength{\bibsep}{1pt}
\begin{small}

\end{small}

\end{document}